\documentclass[aps,prb,twocolumn,preprintnumbers,superscriptaddress,amsmath,amssymb]{revtex4}
\usepackage{bm}
\usepackage[english]{babel}
\usepackage[cp1251]{inputenc}
\usepackage[mathlines]{lineno}
\usepackage{color}
\usepackage{graphicx}

\begin{document}

\title{Tuning of Hybrid Oligomers via Nanoscale fs-Laser Reshaping}

\author{Sergey~Lepeshov}
\affiliation{Laboratory of Nanophotonics and Metamaterials, ITMO University, St.~Petersburg, Russia}

\author{Alexander~Krasnok}
\affiliation{Laboratory of Nanophotonics and Metamaterials, ITMO University, St.~Petersburg, Russia}
\affiliation{Department of Electrical and Computer Engineering, The University of Texas at Austin, Austin, Texas 78712, USA}
\email{krasnokfiz@mail.ru}

\author{Ivan~Mukhin}
\affiliation{Laboratory of Nanophotonics and Metamaterials, ITMO University, St.~Petersburg, Russia}
\affiliation{Laboratory of Renewable Energy Sources, St. Petersburg Academic University, St.~Petersburg, Russia}

\author{Dmitry~Zuev}
\affiliation{Laboratory of Nanophotonics and Metamaterials, ITMO University, St.~Petersburg, Russia}

\author{Alexander~Gudovskikh}
\affiliation{Laboratory of Renewable Energy Sources, St. Petersburg Academic University, St.~Petersburg, Russia}

\author{Valentin~Milichko}
\affiliation{Laboratory of Nanophotonics and Metamaterials, ITMO University, St.~Petersburg, Russia}

\author{Pavel~Belov}
\affiliation{Laboratory of Nanophotonics and Metamaterials, ITMO University, St.~Petersburg, Russia}

\author{Andrey~Miroshnichenko}
\affiliation{Nonlinear Physics Centre, Australian National University, Canberra ACT 2601, Australia}



\begin{abstract}
Various clusters of metallic or dielectric nanoparticles can exhibit sharp Fano resonances originating from at least two modes interference of different spectral width. However, for practical applications such as biosensing or nonlinear nanophotonics, the fine-tuning of the Fano resonances is generally required. Here, we propose and demonstrate a novel type of hybrid oligomers consisting of asymmetric metal-dielectric (Au/Si) nanoparticles with a sharp Fano resonance in visible range, which has a predominantly magnetic origin. We demonstrate both, numerically and experimentally, that such hybrid nanoparticle oligomers allow fine-tuning of the Fano resonance via fs-laser induced melting of Au nanoparticles at the nanometer scale. We show that the Fano resonance wavelength can be changed by fs-laser reshaping very precisely (within 15~nm) being accompanied by a reconfiguration of its profile.
\end{abstract}

\maketitle

\section*{Introduction}
Plasmonic oligomers consisting of nanoparticles of nobel metals (e.g. silver and gold) are the cornerstone of modern nanophotonics due to a sharp effect of resonant scattering originating from destructive interference between super-radiant and sub-radiant modes~\cite{Hentschel2010, Mirin2009, Hillenbrand_NL2011, Halas_NL_2012, Dorpe2012, Koenderink_PRL_2012, Martin2013}, which can be described in terms of the \textit{Fano resonances}~\cite{Lukyanchuk2010, Miroshnichenko2010}. In addition to a strong local field enhancement, the asymmetric profile of the Fano resonance in such structures allows to control the radiative damping of the localized surface plasmon resonance. This superior feature is very useful for applications of nanophotonics, although such plasmonic nanostructures suffer from high dissipative losses in visible~\cite{Boltasseva2011}. Recently, all-dielectric oligomers based on high-index dielectric and semiconductor nanoparticles (e.g. silicon) have been proposed theoretically~\cite{Miroshnichenko_2012, Hopkins2013, Hopkins2013a, Hopkins2015}, and realized experimentally~\cite{Chong2014, Filonov2014, Shcherbakov2015} as a more efficient counterpart to the plasmonic ones. It has been shown that the all-dielectric oligomers can exhibit not only an electric type of Fano resonance, but also a \textit{magnetic one}, which is associated with the optically induced magnetic dipole mode of individual high-index nanoparticle~\cite{Evlyukhin:NL:2012, Kuznetsov:SR:2012, Kuznetsov2016}. An existence of the resonant magnetic response in such structures makes it possible to control the electric and magnetic response individually. It is worth noting that the plasmonic oligomers also can provide resonant magnetic response~\cite{Monticone_2013, Haran_2013, Sun_2016} including more complicated metal-insulator-metal structures~\cite{Hong_2012, Verre2015}, where the insulator has a low refractive index (SiO$_2$). However, such resonant plasmonic structures suffer from dissipative losses inherent to metals in the visible range.

\begin{figure*}
\centering
\includegraphics[width=0.99\textwidth]{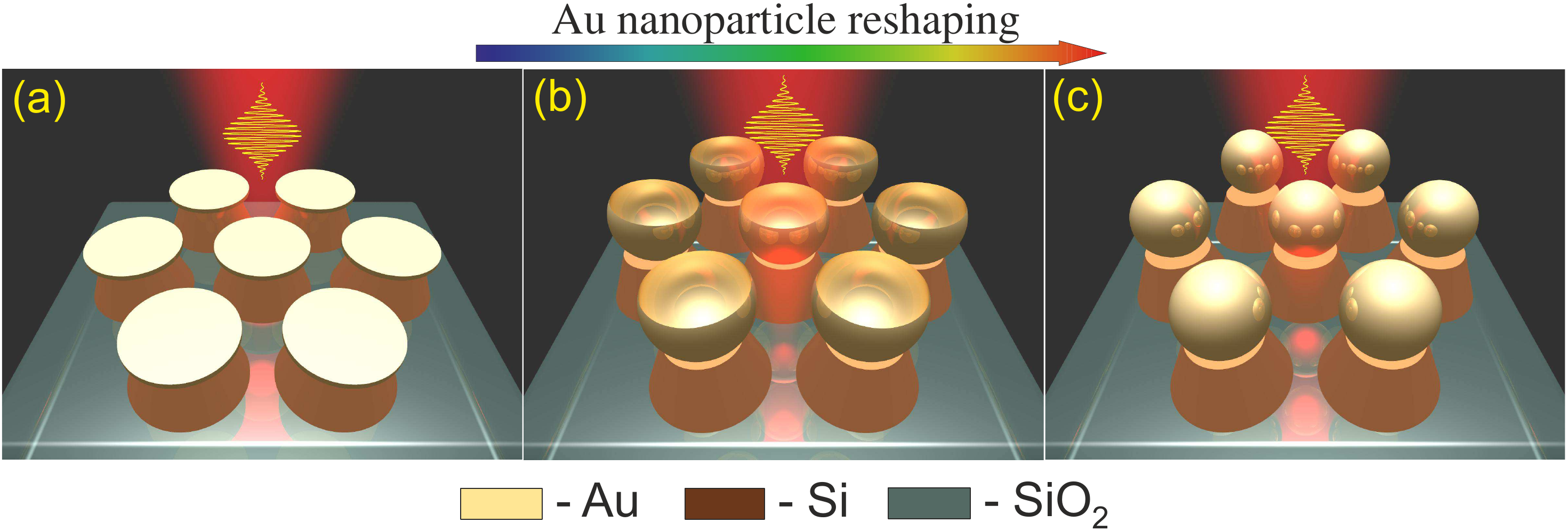}
\caption{Sketch of the hybrid oligomers composed of asymmetric hybrid (Au/Si)~dimer nanoparticles with different shapes of Au components, which correspond to different stages of laser reshaping: (a)~nanodiscs, (b)~nanocups, and (c)~nanospheres.}\label{artistic}
\end{figure*}

Currently, both all-dielectric and more sophisticated plasmonic oligomers are used to achieve the near-field enhancement~\cite{Toma_2015} and associated nonlinear optical effects~\cite{Bragas_2014, Martin_2015, Shorohov2016}, biosensing~\cite{Jin2012, Deng2015}, surface-enhanced Raman scattering~\cite{Dorpe2012}, graphene electronics~\cite{Fang2012}, strong optical activity~\cite{Fang_2016}, which potentially can be applied for quantum optics~\cite{Talebi2012} as well. In terms of these practical applications, it is necessary to have a possibility for a fine-tuning of the spectral features of the Fano resonances in the \textit{fabricated} nanoparticle oligomer structures. The recently proposed approaches to tune the Fano resonances in the clusters of metallic nanoparticles are based on a changing of their geometry during a fabrication process~\cite{GiessenACS2011, Chong2014, Sun2014, King2015} or electromagnetic properties of their environment~\cite{Lassiter2010, Park_2012}. Moreover, the near-field distribution and absorption properties of oligomers with rotational symmetry can be tuned via a polarisation of incident light, leaving the scattering properties unchanged~\cite{Rahmani2013}. Although these methods show significant performance, they can not be applied to fabricated oligomers for fine-tuning of their modes and scattering properties.

The purpose of this paper is twofold. First, we propose to combine two paradigms of plasmonic and all-dielectric oligomers and form a hybrid metal-dielectric clusters to have benefits and advantages of both of them. Recently, unique properties of asymmetric hybrid nanoparticles made them as a very promising platform for nanophotonics~\cite{Henzie2006, Halas_NL2012, Jiang2014, Li2014, Zhu2015, Wang2015, Narasimhan2015}. However, oligomers based on resonant plasmonic and dielectric nanoparticles have not been studied yet. Here, we suggest and implement a novel type of oligomers consisting of resonant asymmetric metal-dielectric (Au/Si) nanoparticles realizing the concept of hybrid oligomers. We show that the proposed oligomers exhibit a sharp Fano resonance in the visible range. Based on the multipole expansion analysis (for the method of multipole expansion in vector spherical harmonics see \textit{Supplementary Information}), we demonstrate that the Fano resonance has a predominantly magnetic origin owing to magnetic Mie-type modes of the Si nanoparticles. Second, being inspired by our recent experimental work~\cite{Zuev_2016} on a new technique for fabrication of asymmetric hybrid (Au/Si) nanoparticles, we propose and realize an original approach for tuning of the magnetic Fano resonance in the oligomers. The approach is based on a fs-laser induced melting of Au part of hybrid dimer nanoparticles at the nanometer scale (as schematically shown in Figure~\ref{artistic}). We show that the Fano resonance wavelength can be changed by fs-laser reshaping very precisely being accompanied by a reconfiguration of its profile.

\begin{figure*}
	\centering
	\includegraphics[width=0.99\textwidth]{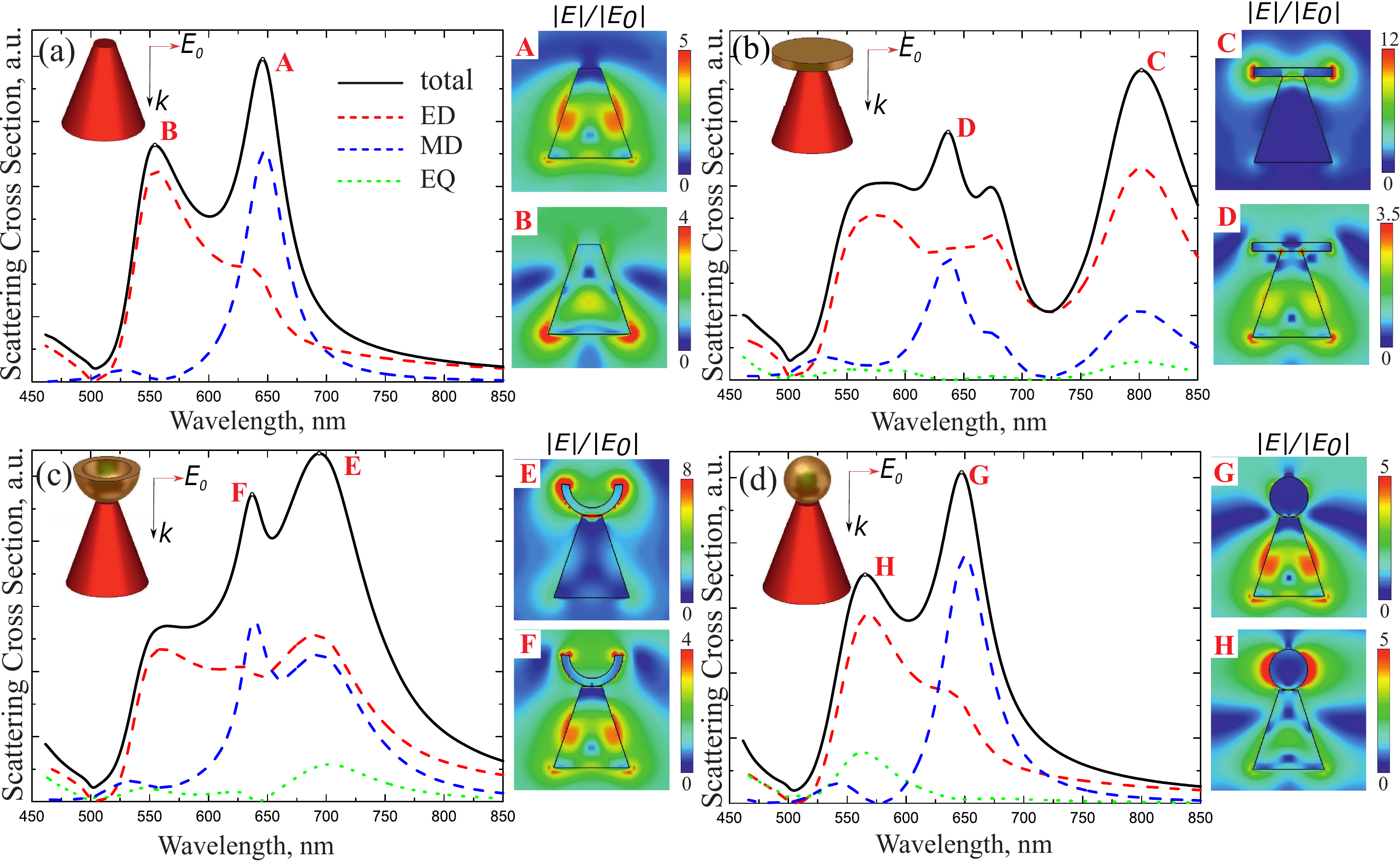}
	\caption{The scattering cross sections (black curves) and results of the multipole expansion for (a)~a single Si nanocone, and hybrid Au/Si nanoparticles: Si nanocone with (b)~Au nanodisc, (c)~Au nanocup, and (d) Au nanosphere. The diameters of the lower base of the Si nanocone and the Au nanodisc are equal 190~nm. The spectra are normalised identically. (A--H)~The electric field profiles (in terms of field amplitude) at the corresponding resonance; the corresponding points are marked on the spectra. The incident wave propagates along the axis of symmetry of the nanoparticles.}\label{unit}
\end{figure*}

\section*{Results}
We start our analysis by comparing the scattering cross sections of a single silicon nanocone and hybrid nanoparticles (see Figure~\ref{unit}). We assume the incident planewave is propagating along the axis of symmetry of the nanoparticles. Figure~\ref{unit}(a) shows the light scattering spectrum of the single Si truncated nanocone. The geometric parameters of the cone are taken from the Ref.~\cite{Zuev_2016}, namely, the diameter of the upper base is $a = 60$~nm, and the cone height is $h = 200$~nm. The lower base ($b$) of the Si nanocone is $b=190$~nm; the results for Si nanocones of other sizes are presented in \textit{Supplementary Information}.

The scattering cross section of the single Si nanocone has two distinct resonances, around the wavelengths 550~nm and 650~nm (Figure~\ref{unit}(a), points A and B). By using the multipole expansion~\cite{Jackson1999, Grahn2012}, we reveal that these resonances are of the electric dipole type at $550$ nm (ED, red dashed curve) and magnetic dipole type at $650$ nm (MD, blue dashed curve). Higher-order multipoles produce a negligible contribution for given parameters. The electric near-field distribution profiles at these resonances are presented in Figure~\ref{unit}(a)A,B. It is known that the magnetic dipole Mie resonance condition for the dielectric (such as silicon) nanoparticle depends on its size. For the conical particle under investigation we obtain the following equation for the wavelength of magnetic resonance $\lambda_{\rm res}\approx 0.9 b n_{\rm d}$, where $\lambda_{\rm res}$ is the resonant wavelength, $n_{\rm d}$ is the refractive index of the silicon nanoparticle~\cite{Vuye_1993}. In previous experimental articles it has been shown that the refractive index of balk crystalline silicon works well with nanoparticles of such sizes~\cite{Evlyukhin:NL:2012, Kuznetsov:SR:2012, Dmitriev2016, Dmitriev_NL_2016}. The last equation is a good approximation in the domain close to selected geometric parameters. Thus, the reduction of $b$ from 190~nm to 150~nm leads to the blue-shift of the magnetic resonance from $\lambda_{\rm res}=640$~nm to $\lambda_{\rm res}=570$~nm (see \textit{Supplementary Information}). This feature was recently used for controlling over the wavelength of Fano resonance in all-dielectric oligomers at the manufacturing stage~\cite{Chong2014}.

Now we consider the scattering properties of a single hybrid nanoparticle consisting of Si nanocone and Au nanodisc (see inset in Figure~\ref{unit}(b)). We assume that the diameter of Au nanodisc is equal to the diameter of the lower base of the Si nanocone, which is dictated by the lithography process~\cite{Zuev_2016}. We also take the thickness of the Au nanodisc is equal to $d = 20$~nm. By adding the gold nanodisc on the upper base of Si nanocone, an additional resonance appears in the scattering spectra of the resulting hybrid nanoparticle, which is shown in Figure~\ref{unit}(b), and where the resonance depicted by point $C$. This resonance has a plasmonic nature and manifests itself in strong local electric field enhancement around the nanodisc. Moreover, the modes of Si nanocone and Au nanodisc begin to hybridize. The hybridization of Mie and plasmonic modes causes their mutual perturbation (see multipole expansion for this particle in Figure~\ref{unit}(b)). The magnetic Mie resonance still has a resonant behaviour (Figure~\ref{unit}(b), point $D$). The electric near-field distribution at the wavelength of the plasmonic resonance ($\lambda=800$~nm) is presented in Figure~\ref{unit}(b)C. The existence of the Au nanodisc perturbs the electric near-field of the nanocone at its magnetic resonance (Figure~\ref{unit}(b)D) due to their effective coupling. The scattering properties of the hybrid nanoparticles in the optical frequency range and electric near-field distributions hereinafter were numerically calculated by using CST~Microwave~Studio. A nonuniform mesh was used to improve the accuracy in the vicinity of the Au nanoparticle where the field concentration was significantly large. The dispersion model for the Au and Si materials was taken from the literature~\cite{Palik_1985, Meyer_2006, Christy_1972, Vuye_1993}.

The plasmon resonance of Au nanoparticles arises from an excitation of localized surface plasmon modes, which are strongly depended on the geometrical shape of the nanoparticle~\cite{Giannini2011, Zhu2015, Viarbitskaya2015, Makarov2016}. It has been shown that under irradiation of a Au nanoparticle by femtosecond laser pulse with energy density of 40--50~mJ/cm$^2$ (depending on the Au particle size), the Au nanoparticle changes its shape from a disc to a cup~\cite{Halas_2008, HalasNL_2011, Zuev_2016}. At lower intensities, there is no detectable shape deformation. We emphasise that it is necessary to use a truncated nanocone to properly change the Au nanoparticles shape. At the same time, the Si nanocone is not affected by the fs-laser radiation due to the higher melting temperature and enthalpy of fusion (about 1687~K and 50.21 kJ/mol for crystalline silicon in contrast to 1337~K and 12.55 kJ/mol for gold). The plasmon resonance of the deformed nanoparticle [see scattering spectra in Figure~\ref{unit}(c)] shifts to shorter wavelengths (from 800~nm to 690~nm, in our case). Now it is difficult to separate the response of whole hybrid nanoparticle to responses of dielectric and metallic parts. It results in dramatically changing in the near field distribution of the hybrid nanoparticle [see Figure~\ref{unit}(c)E] and appearance of hot-spots of the locally enhanced electric field at the edges of the nanocup where $E/E_{0}$ reaches 8, $E_{0}$ is the exciting field strength. Upon the Au nanoparticle reshaping, the wavelength of the magnetic resonance of Si nanocone shifts to 630~nm (see Figure~\ref{unit}(c)F). Moreover, in this case we observe the notable contribution of the electric quadrupole mode (EQ) in the total scattering (see Figure~\ref{unit}(c), green doted curve).

By increasing the energy density of the laser radiation up to 70--80~mJ/cm$^2$, the nanocup transforms its shape to a nanosphere (in our case the radius of resulting sphere is 51~nm). The scattering cross section of such hybrid nanoparticle as well as the results of multipole expansion are presented in Figure~\ref{unit}(d). The scattering cross section is similar to the single Si nanocone, due to the Au nanosphere scatters much less of light energy than the Si nanocone. Thus, the position of the Au nanoparticle plasmon resonance as well as response of the whole hybrid nanoparticle can be controlled via fs-laser induced reshaping.

\begin{figure*}
\centering
\includegraphics[width=0.99\textwidth]{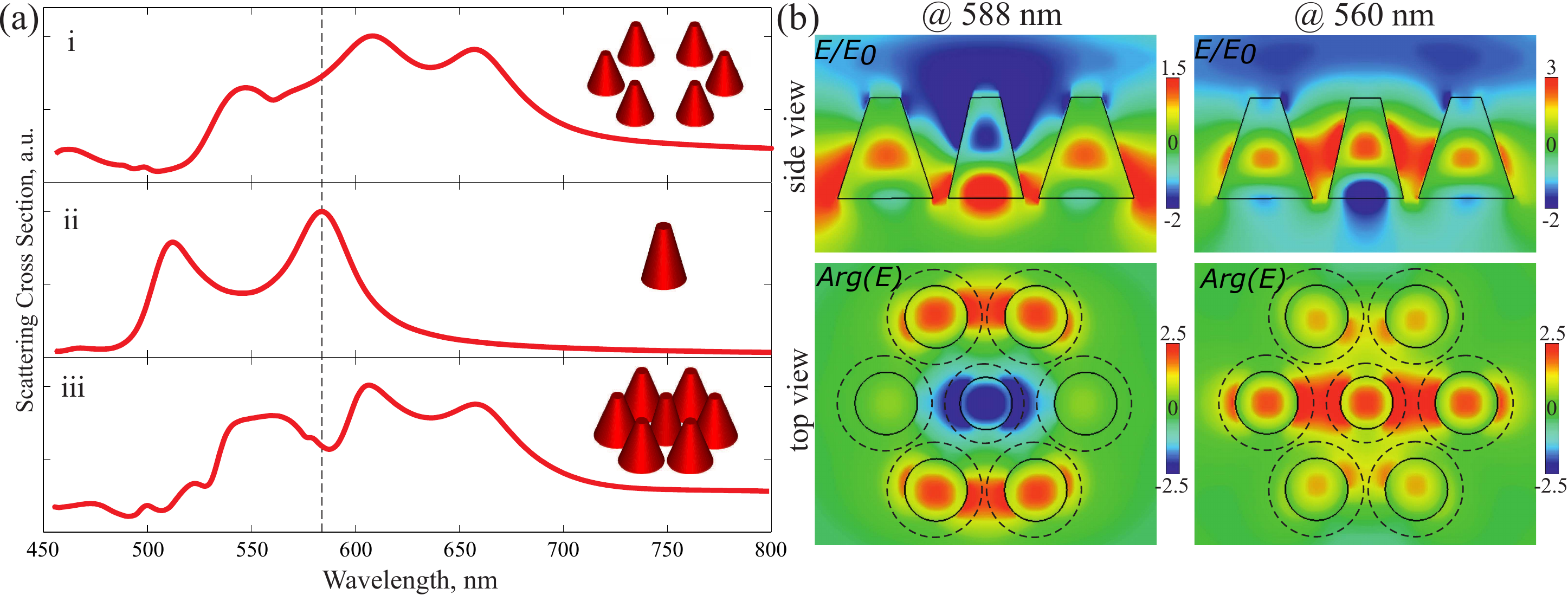}
\caption{(a)~Scattering cross sections of (i)~all-dielectric hexamer based on Si nanocones, (ii)~single Si nanocone with smaller lower base, and (iii)~all-dielectric heptamer. The gaps between the central cone and the boundary ones are 30~nm. The dashed line shows the position of the Fano resonance. (b)~The electric field profiles (in terms of field amplitude) in the vertical and top cross-sections calculated at the scattering intensity dip of Fano resonance ($\lambda=588$~nm) and outside of the resonance ($\lambda=560$~nm).}\label{cones1}
\end{figure*}

Let us consider an all-dielectric oligomer consisting of Si nanocones and having a 6-fold rotational axis ($C_6$). To demonstrate that the oligomer has a Fano resonance, we calculate the scattering spectra of hexamer and single Si nanocone separately as well as scattering spectra of whole oligomer (see Figure~\ref{cones1}). The hexamer structure is based on the nanocones with a diameter of the lower base of $b=190$~nm. The gap between nanocones (the distances between the neighboring lower bases) is 10~nm, which leads to their effective interaction resulting in the appearance of low-Q collective modes. The scattering spectra of these modes overlap forming a non-resonant scattering channel [see Figure~\ref{cones1}(a)i]. To obtain a heptamer, we place a Si nanocone with the diameter of lower base of 150~nm and with a relatively narrow magnetic resonance (see Figure~\ref{cones1}(a)ii) in the center of the hexamer. The gap between the central Si nanocone and the hexamer's ones in the resulting structure is 30~nm. Figure~\ref{cones1}(a)iii shows the scattering spectrum of the resulting heptamer. This spectrum has a resonant dip at the wavelength of magnetic Mie resonance of the central nanocone (around $\lambda=590$~nm) with a pronounced asymmetric profile. In Refs.~\cite{Miroshnichenko_2012, Chong2014} it has been shown that this dip is associated with the \textit{magnetic Fano resonance}, which is caused by the scattered wave interference of two modes -- the spectrally narrow magnetic dipole Mie mode of the central nanocone and the broadband collective magnetic modes of the hexamer. The Fano resonance dip at 588~nm is caused by antiphase oscillating of the magnetic dipoles of heptamer and magnetic dipole of the central nanocone (dark mode). Outside of this resonance ($\lambda=560$~nm) these modes oscillate in phase (bright mode) (see Figure~\ref{cones1}(b)). We also note that due to the rotational symmetry of the all-dielectric oligomer the scattering cross section does not depend on the incident wave polarization~\cite{Hopkins2013, Fuller_1994, Miroshnichenko_ACSN_2013}. The shape of the Fano resonance depends on the distances between the nanocones. The results of detailed study of this effect are presented in \textit{Supplementary Information}.

\begin{figure*}
\centering
\includegraphics[width=0.99\textwidth]{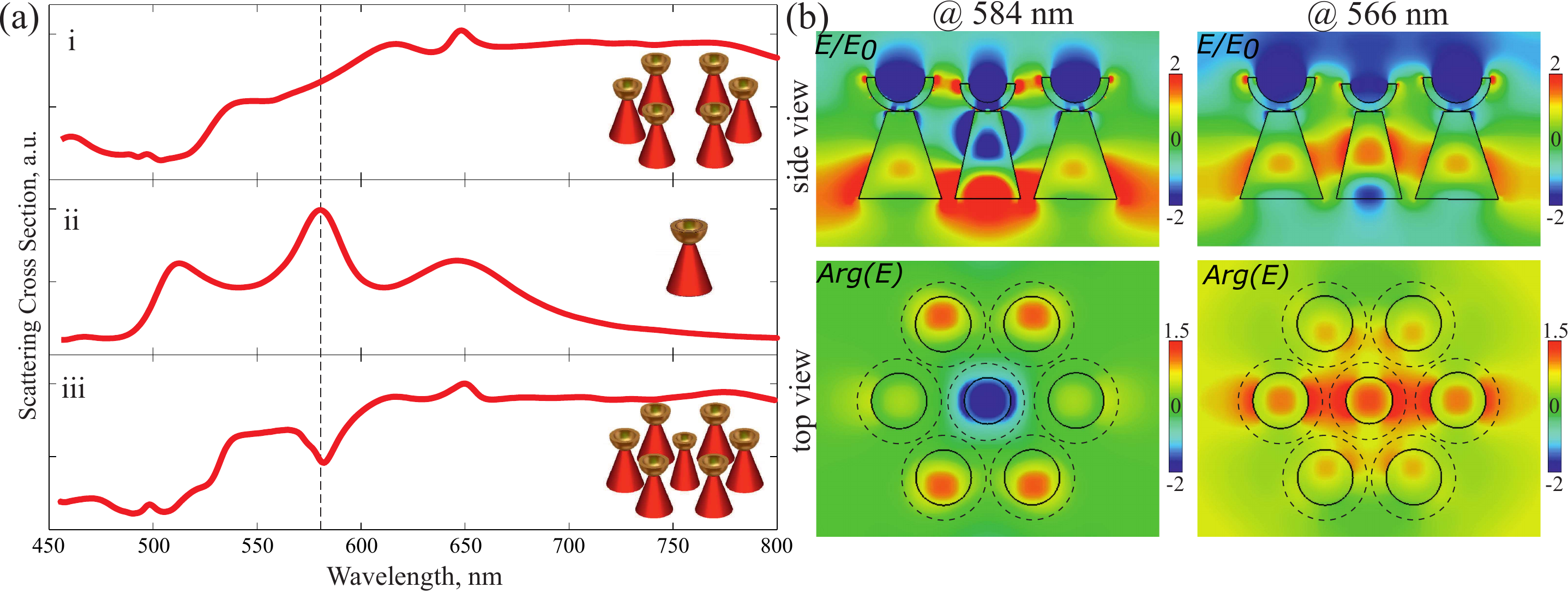}
\caption{(a)~Scattering cross sections of (i)~hybrid Au/Si hexamer, (ii)~single hybrid nanoparticle with smaller lower base, and (iii)~heptamer, with Au nanoparticles in form of nanocups. (b)~The electric field profiles (in terms of field amplitude) in the vertical and top cross-sections calculated at the scattering intensity dip of Fano resonance ($\lambda=584$~nm) and outside of the resonance ($\lambda=566$~nm).}\label{caps}
\end{figure*}

Our goal now is to use the melting of the Au nanoparticles placed on the Si nanocones to tune the magnetic Fano resonance in the hybrid oligomer. For this propose we show that the hybrid oligomer has a pronounced Fano resonance, even in the presence of Au nanocups, i.e. when the Au nanoparticle is resonant in vicinity of the Fano resonance wavelength. In other words, we show that the Au nanoparticles perturb the Fano resonance, but does not destroy it. We demonstrate it in the same manner as for the all-dielectric oligomers. Namely, the scattering spectra of the the hybrid Au/Si hexamer with Au nanocups has a broad and nonresonant wing of collective modes (see Figure~\ref{caps}(a)i). The interaction of these modes with the narrow resonance of the single Au/Si nanoparticle (see Figure~\ref{caps}(a)ii) results in appearance of the asymmetrical dip in the scattering spectrum (see Figure~\ref{caps}(a)iii). The electric field distribution profiles in the side and top views calculated at the scattering intensity dip of Fano resonance ($\lambda=584$~nm) and outside of the resonance ($\lambda=566$~nm) are presented in Figure~\ref{caps}(b). At the Fano resonance wavelength ($\lambda=584$~nm) the modes of central particle and hexamer oscillate in opposite phase, forming a dark mode of the whole hybrid oligomer.

\begin{figure*}
\centering
\includegraphics[width=0.99\textwidth]{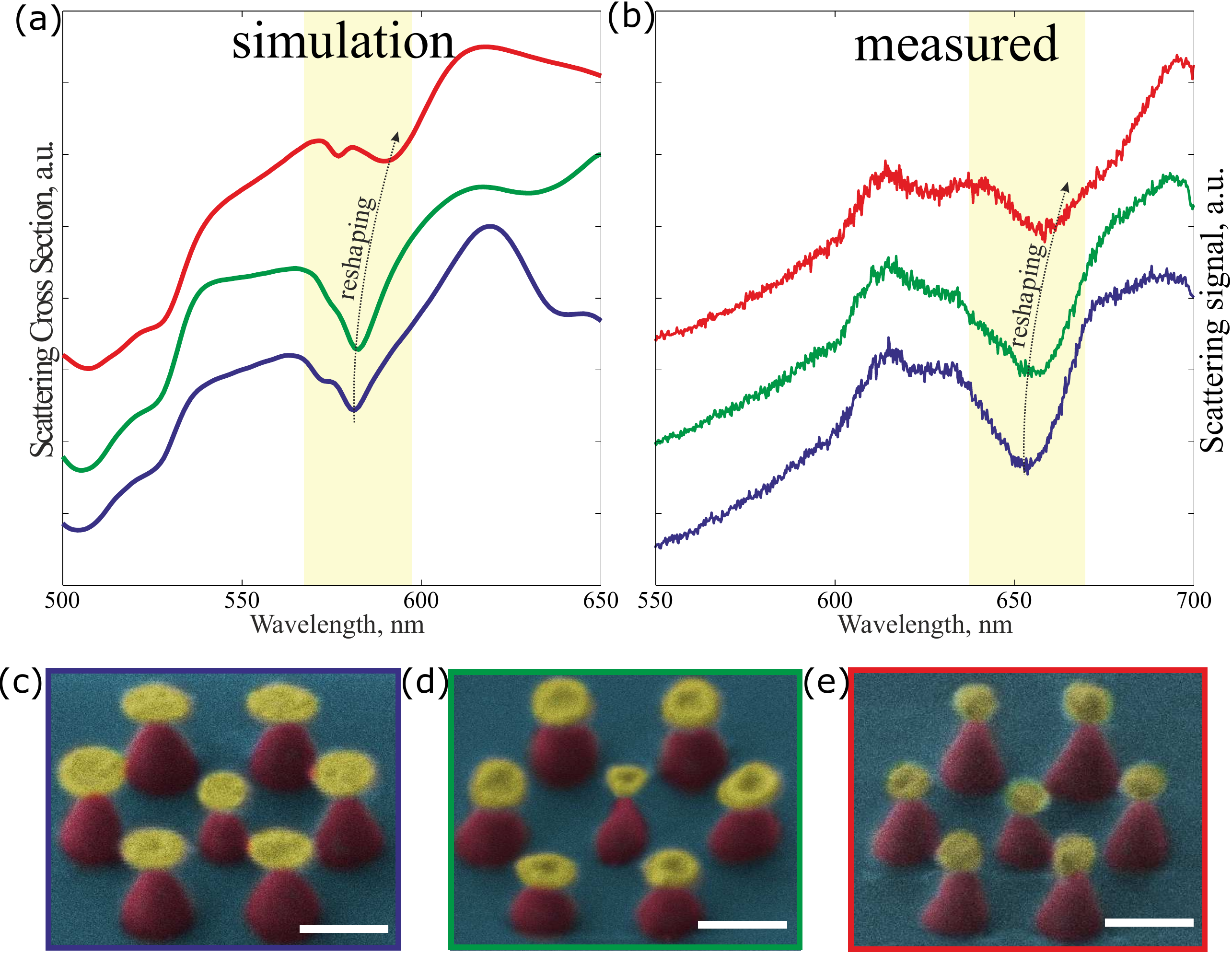}
\caption{(a)~Calculated scattering cross section spectra and (b)~experimentally measured scattering dark-field signals of the hybrid Au/Si heptamer with Au nanodiscs (blue curve), Au nanocups (green curve), and Au nanospheres (red curve).The spectral region with strong Fano resonane response is highlighted by yellow stripe. (c)--(e)~The SEM images (viewing angle is 45$^\circ$) of typical of fabricated hybrid oligomers with Au nanodiscs~(c), Au nanocups~(d), and Au nanospheres~(e); the scale-bar is 200~nm.}\label{hybrid}
\end{figure*}

Now we consider the optical properties of hybrid oligomers composed of Au/Si nanoparticles for different stages of reshaping (see Figure~\ref{hybrid}). We study the hybrid oligomers with the diameters of the lower base of the hexamer's Si nanocones and the central nanocone of $190$~nm and 150~nm, respectively. The gap between the central Si nanocone and the hexamer's ones is 30~nm. We consider the Fano resonance of the oligomer composed of hybrid nanoparticles with Au nanodiscs that appears at 580~nm (see Figure~\ref{hybrid}(a), blue curve). The resonance is caused mainly by the responses of Si nanocones, and its wavelength corresponds to the wavelength of the Fano resonance of all-dielectric oligomer (see Figure~\ref{cones1}(a)iii) because of the weak coupling between Au nanodiscs and Si nanocones. For this case of Au/Si nanoparticle oligomers the Fano resonance is less pronounced compared to the all-dielectric counterpart (see Figure~\ref{cones1}(a)).

Next, we numerically show that the profile and its spectral position of this resonance can be changed by fs-laser induced reshaping of the Au nanodiscs. When the Au nanoparticles take the form of nanocups, the minimum of the Fano resonance shifts to $\lambda=585$~nm being accompanied by a constriction of its profile (see Figure~\ref{hybrid}(a), green curve). It has been shown above (see Figure~\ref{caps}(a)) that this very pronounced dip in the scattering spectrum of hybrid oligomers with Au nanocups corresponds to the Fano resonance. Upon further reshaping of the Au nanoparticles to nanospheres, the Fano resonance becomes broader again and its minimum shifts to $\lambda=595$~nm (see Figure~\ref{hybrid}(a), red curve). Thus, the laser reshaping of the Au nanoparticles can be applied for fine-tuning of the Fano resonance of the hybrid oligomers.

In order to prove the concept of fine-tuning of the magnetic Fano resonance we provide a series of dark-field scattering spectra measurements from hybrid oligomers with different degree of the Au nanoparticle reshaping. First, we have fabricated the hybrid oligomers based on the gold nanodiscs placed on the upper base of the truncated silicon nanocones by means of a combination of e-beam lithography (25~kV), metal evaporation, lift-off procedure, and gas-phase chemical etching. Recently, it has been shown that under the electron-beam processing step with 25~kV acceleration voltage the amorphous silicon gets the nanocrystalline structure~\cite{Baranov_ACSPh_2016}. This method of hybrid nanostructures fabrication is developed in Ref.~\cite{Zuev_2016}. At the first step, an a-Si:H layer with a thickness of $\approx$200~nm was deposited on a properly cleaned substrate of fused silica by the plasma-enhanced chemical vapor deposition of SiH$_3$ gas. Then, the arrays of metal nanodiscs consisting of Cr/Au layers with thicknesses of $\approx$ 1~nm/20~nm were produced by means of e-beam lithography, metal deposition, followed by the lift-off procedures. After that, the silicon layer was etched through a fabricated metal mask using radio-frequency inductively coupled plasma technology in the presence of SF$_6$ and O$_2$ gases. The etching was carried out with temperature of 265~K to fabricate Si nanostructures in the shape of nanocones. The typical SEM images of fabricated hybrid oligomers with Au nanodiscs, Au nanocups, and Au nanospheres are represented in Figure~\ref{hybrid}(c)--(e); the scale-bar is 200~nm.

For fs-laser melting a commercial femtosecond laser system (Ytterbium-doped Femtosecond Solid-State Laser TeMa 150, Avesta~Poject) was used, providing laser pulses at a central wavelength of 1050~nm, with a maximum pulse energy of 85~nJ, and a pulse duration of 150~fs at a repetition rate of 80~MHz. The laser energy was varied and controlled by an optical attenuator (Avesta~Poject) and a power meter (FielfMax~II, Coherent), respectively. Laser pulses were focused on the fabricated sample by an objective (Mitutoyo~M~Plan~Apo~NIR~10X) with a numerical aperture (NA) of 0.26. In order to adhesion of the Au nanodisk and Si nanocone a thin Cr layer is used. According to the results of molecular dynamic simulations~\cite{Zuev_2016}, the Cr layer (with a thickness of 1-2~nm) provides a desired shape of the Au nanoparticle during laser reshaping without affecting the electromagnetic properties of the hybrid nanoparticle. Moreover, this Cr layer prevents formation of the Au-Si alloy.

Measurements of the scattering spectra were carried out in a dark-field scheme, where the arrays irradiation was performed by p-polarized light from a halogen lamp (HL-2000-FHSA) at an angle of incidence of 70$^\circ$ with the surface normal. Scattered signal collection was performed by means of a Mitutoyo~M~Plan~APO~NIR objective (NA = 0.7), which directed light to a commercial spectrometer (Horiba~LabRam~HR) with a CCD camera (Andor~DU~420A-OE~325). The confocal optical scheme was optimized for signal collection from individual nanoparticles. A sketch of the experimental setup for the polarization-resolved dark-field spectroscopy is represented in the \textit{Supplementary Information}.

The results of fine-tuning of the Fano resonance in the fabricated hybrid oligomers are summarized in Figure~\ref{hybrid}(b). Our experimental results clearly show the spectral shift of the Fano resonance minimum from $\lambda=650$~nm to $\lambda=660$~nm with increasing of the power of external laser field from 0~mW (blue curve) to 40~mW (green curve). Following our previous results in this regime of reshaping the Au nanoparticles takes the form of nanocones~\cite{Zuev_2016}. Upon further increase of the laser power up to 90~mW the Au nanocones reshape to nanospheres with spectral shifting of the Fano resonance dip to $\lambda=665$~nm (see Figure~\ref{hybrid}(b), red curve). The measured damage threshold of the Au nanoparticles is about 130~mW. At this power the Fano resonance of the hybrid oligomer disappears. The slight mismatching of the numerical and experimental results is explained by the presence of SiO$_2$ substrate and accuracy of nanostructures fabrication.

\section*{Conclusion}
In summary, we have proposed and implemented a novel type of hybrid oligomers consisting of resonant asymmetric metal-dielectric (Au/Si) nanoparticles and exhibiting a sharp Fano resonance in visible range, which has a predominantly magnetic nature. We have demonstrated, numerically and experimentally, that such oligomers allow irreversible fine-tuning of the Fano resonance via fs-laser melting of Au nanoparticles at the nanometer scale. We have shown that the Fano resonance wavelength can be changed by fs-laser reshaping very precisely (within 15~nm) being accompanied by a reconfiguration of its profile. We believe that our results pave the way to realization of nanophotonic elements that can be adjusted after their manufacturing.

\section*{Acknowledgements}
This work was financially supported by Russian Science Foundation (Grant 15-19-30023).

The authors declare no competing financial interest.


\end{document}